\documentclass[8.5pt,twoside,twocolumn]{article}
\usepackage[super,sort&compress,comma]{natbib}
\usepackage[version=4]{mhchem}
\usepackage{times,mathptmx}
\usepackage{sectsty}
\usepackage{balance}
\usepackage{graphicx} 
\usepackage{lastpage}
\usepackage[format=plain,justification=raggedright,singlelinecheck=false,font=small,labelfont=bf,labelsep=space]{caption}
\usepackage{fancyhdr}
\usepackage{graphicx}
\usepackage{dcolumn}
\usepackage{bm}
\usepackage{color}
\usepackage{siunitx}

\usepackage{xcolor}

\usepackage{tikz}
\usetikzlibrary{calc}
\begin{document}
\thispagestyle{plain}
\fancypagestyle{plain}{
\fancyhead[L]{\includegraphics[height=8pt]{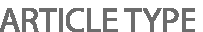}}
\fancyhead[C]{\hspace{-1cm}\includegraphics[height=20pt]{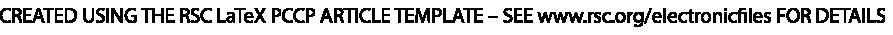}}
\fancyhead[R]{\includegraphics[height=10pt]{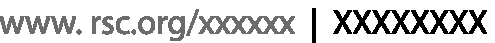}\vspace{-0.2cm}}
\renewcommand{\headrulewidth}{1pt}}
\renewcommand{\thefootnote}{\fnsymbol{footnote}}
\renewcommand\footnoterule{\vspace*{1pt}%
\hrule width 3.4in height 0.4pt \vspace*{5pt}}
\setcounter{secnumdepth}{5}

\makeatletter
\def\subsubsection{\@startsection{subsubsection}{3}{10pt}{-1.25ex plus -1ex minus -.1ex}{0ex plus 0ex}{\normalsize\bf}}
\def\paragraph{\@startsection{paragraph}{4}{10pt}{-1.25ex plus -1ex minus -.1ex}{0ex plus 0ex}{\normalsize\textit}}
\renewcommand\@biblabel[1]{#1}
\renewcommand\@makefntext[1]%
{\noindent\makebox[0pt][r]{\@thefnmark\,}#1}
\makeatother
\renewcommand{\figurename}{\small{Fig.}~}
\sectionfont{\large}
\subsectionfont{\normalsize}

\fancyfoot{}
\fancyfoot[LO,RE]{\vspace{-7pt}\includegraphics[height=9pt]{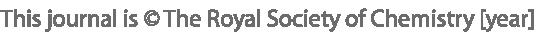}}
\fancyfoot[CO]{\vspace{-7.2pt}\hspace{12.2cm}\includegraphics{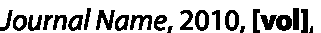}}
\fancyfoot[CE]{\vspace{-7.5pt}\hspace{-13.5cm}\includegraphics{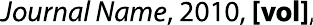}}
\fancyfoot[RO]{\footnotesize{\sffamily{1--\pageref{LastPage} ~\textbar  \hspace{2pt}\thepage}}}
\fancyfoot[LE]{\footnotesize{\sffamily{\thepage~\textbar\hspace{3.45cm} 1--\pageref{LastPage}}}}
\fancyhead{}
\renewcommand{\headrulewidth}{1pt}
\renewcommand{\footrulewidth}{1pt}
\setlength{\arrayrulewidth}{1pt}
\setlength{\columnsep}{6.5mm}
\setlength\bibsep{1pt}

\twocolumn[
  \begin{@twocolumnfalse}
\noindent\LARGE{\textbf{Collective hydrodynamic transport of magnetic microrollers$^\dag$}}

\vspace{0.6cm}
\noindent\large{\textbf{Gaspard Junot,\textit{$^{a}$} Andrejs Cebers,\textit{$^{b}$}
and
Pietro Tierno\textit{$^{acd}$}}}\vspace{0.5cm}

\noindent\textit{\small{\textbf{Received Xth XXXXXXXXXX 20XX, Accepted Xth XXXXXXXXX 20XX\newline
First published on the web Xth XXXXXXXXXX 200X}}}

\noindent \textbf{\small{DOI: xxxxx}}
\vspace{0.6cm}

\noindent \normalsize{
We investigate the collective transport properties of microscopic magnetic rollers that 
propel close to a surface due to a circularly polarized, rotating magnetic field. The applied field exerts a torque to the particles, which induces a net rolling motion close to a surface. The collective dynamics of the particles result from the balance between magnetic dipolar interactions and hydrodynamic ones. We show that, when hydrodynamics dominate, i.e. for high particle spinning, the collective mean velocity linearly increases with the particle density. In this regime we analyse the clustering kinetics, and find that 
hydrodynamic interactions between the anisotropic, elongated particles, induce preferential cluster growth along a direction perpendicular to the driving one, leading to dynamic clusters that easily break and reform during propulsion. 
}
\vspace{0.5cm}
 \end{@twocolumnfalse}
]

\section{Introduction}
\footnotetext{\dag~Electronic Supplementary Information (ESI) available:
Three .WMV videos showing the
dynamics of the hematite microroller at different frequencies. See DOI: xxxx/b000000x/}
\footnotetext{\textit{$^{a}$~Departament de F\'isica de la Mat\`eria Condensada, Universitat de Barcelona, Barcelona, Spain E-mail: ptierno@ub.edu}}
\footnotetext{\textit{$^{b}$~MMML Lab, Department of Physics, University of Latvia, Jelgavas-3, Riga, LV-1004, Latvia.}}
\footnotetext{\textit{$^{c}$~Universitat de Barcelona Institute of Complex Systems (UBICS), Universitat de Barcelona, Barcelona, Spain}}
\footnotetext{\textit{$^{d}$~Institut de Nanoci\`encia i Nanotecnologia, Universitat de Barcelona, 08028, Barcelona, Spain. }}
Microscopic surface rotors driven by a time-dependent magnetic field,
represent a versatile and simple, yet non trivial, model system to 
investigate 
emergent collective phenomena,~\cite{Dobnikar2013,Martin2013,Klapp2016,Koohee2018} similar to  
those observed in ensemble of active and self-propelling particles.~\cite{Mar13,Bec16}
An external, time-dependent field is used to apply a magnetic torque to the particles, and to induce a net spinning close to a substrate.~\cite{Tierno2008,Morimoto2008,Sin10,Pet12,Reenen2015,Tas16,Mai16,Kaiser2017,Koo18}
Since these particles have size within the micrometer range 
and speed of the order of few $\rm{\mu m s^{-1}}$
(similar to that of bacteria or cells), their dynamics occur at effectively 
low Reynolds number ($Re$).
Under these conditions, 
inertia is negligible and Navier-Stokes 
equations become time-reversible.~\cite{Happel1973}
In this situation propulsion is possible only when 
the motion is non-reciprocal, 
namely based on a sequence of body/shape changes that are not identical when reversed in time.~\cite{Purcell1977} Rotating helical flagella in bacteria,~\cite{Berg1973,DiLuzio2005} flexibility in the  tail~\cite{Bibette2005,Gao10} or chemical 
reactions,~\cite{Howse2007,Popescu20} represent some ways to circumvent this constraint. 
In contrast, 
surface rollers present a relative simple 
mechanism of motion, based on the rotational/translational coupling with a close substrate.~\cite{Goldman1967} 
Moreover, magnetic dipolar forces between the microrollers can be tuned by an external field, a feature that allows to investigate  
the role of such forces in the organization process and their interplay with other interactions such as hydrodynamic ones. 

In this context, 
some of us  
recently investigate the dynamics 
of 
pair of propelling microrollers composed of 
anisotropic hematite particles that are
driven close to a plane by an external rotating magnetic field.~\cite{Mar18,Massana2019} 
In such system, hydrodynamic interactions (HIs) between the particles induce the formation of stable bound states, where the particles align tip to tip, namely with 
their relative position parallel
to the long axis, even if dipolar forces are repulsive in this configuration. 
However, the collective effects generated by several interacting microrollers
and how the produced flow field alters the global mean velocity 
remain still unclear. 

In this article we investigate the collective dynamics of interacting microrollers, 
and show that the average translational speed linearly raises with density. 
A similar behaviour was recently reported with different magnetic particles rolling close to a surface~\cite{Driscoll2017}, a fact that enforces its generic 
nature. 
For
our system, this feature cannot be explained by considering
only dipolar forces. Furthermore, this speed up effect can
be controlled by tuning the driving frequency which set the
strength of the hydrodynamic interactions.
We rationalize this observation by calculating
the flow field generated by a set of rotating point particles. 
Further, we analyse the clustering process, and show that 
HIs lead to the formation of dynamic clusters 
which preferentially grow perpendicular to the propulsion direction. 

\section{Experimental details}
\begin{figure}[t]
\begin{center}
\includegraphics[width=\columnwidth]{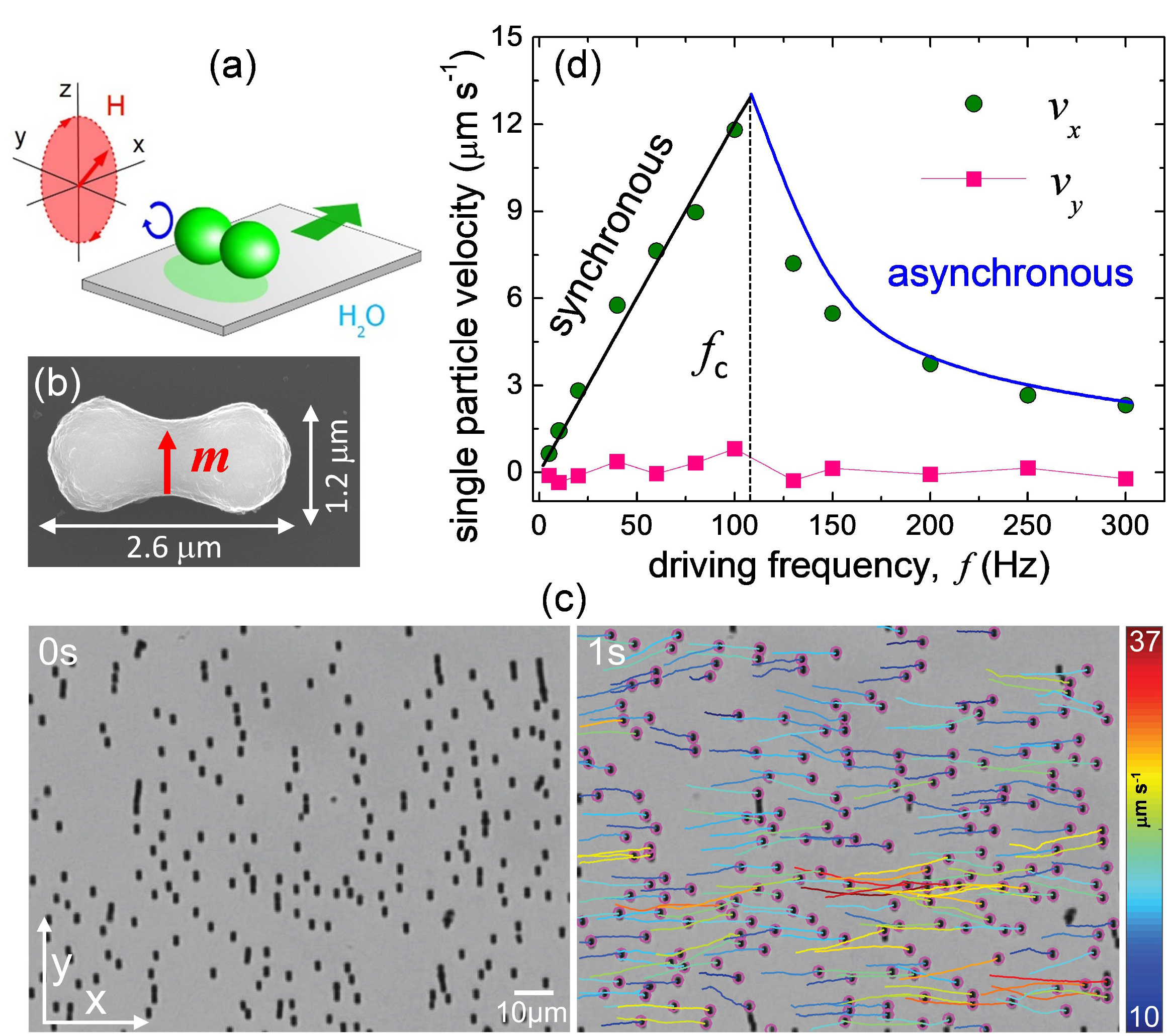}
\caption{(a) Schematic showing a 
hematite particle rolling close to a plane due 
to an external rotating magnetic field 
$\bm {H}$. 
(b) Scanning electron microscope image of one hematite particle 
with the direction of the magnetic moment superimposed in red. 
(c) Sequence of microscope images showing the transport 
of several magnetic particles subjected to a rotating field with 
amplitude $H= 4400 \, \rm{A m^{-1}}$,
and driving frequency $f = 100$Hz.
The particle trajectories over the last second are superimposed 
to the second image.
The colour bar shows the mean particle velocity along
the trajectory.
(d) Single particle velocity along the propulsion direction 
$v_x$ (circles) and perpendicular to it $v_y$ (squares) versus 
driving frequency $f$.
The continuous black and blue lines 
are fits to the synchronous and asynchronous 
regimes, respectively. 
\label{figure1}}
\end{center}
\end{figure}

As shown in Fig.\ref{figure1}, our magnetic microrollers are monodisperse 
hematite particles with a peanut-like shape.
This shape results from the particular fabrication process, more details can be found in a previous work.~\cite{Mar18} 
From the analysis of images obtained with scanning electron microscopy, as shown in the inset in Fig.\ref{figure1}(b),
we find that the hematite particles are characterized by two connected lobes 
with a long (short) axis equal to $\alpha=2.6 \, {\rm \mu m}$ ($\beta=1.2\,  {\rm \mu m}$).
The hematite particles present a small permanent dipole moment oriented 
perpendicular to their long axis. This feature can be explained  
by considering the magnetic
structure of hematite, that crystallizes in the corundum form, 
where Fe cations
are aligned antiferromagnetically.~\cite{Shull1951,Mansilla2002}
The magnetic moment of the hematite ellipsoids was previously determined by measuring the particle relaxation 
under a static field,~\cite{Martinez18} and is 
of the order $m \simeq 9 \cdot 10^{-16} \, \rm{Am^2}$.

After synthesis, the particles are dispersed in pure water (Milli-Q system, Millipore),
with sodium dodecyl sulfate (SDS) at a concentration $1.4 \cdot 10^{-3} \, \rm{g  mL^{-1}}$, and let sediment above a plastic petridish. 
The SDS gives a steric layer to the particles which avoids irreversible particle sticking due to 
attractive Van der Waals interaction.
Further, we raise the pH of the solution between $9$ and $9.5$
by adding Tetramethylammonium Hydroxide. 
The particles remain confined close to the plate due to gravity,
where they float at a distance $h$ there.

The system dynamics are visualized using a light microscope
(Eclipse Ni, Nikon) equipped with a CCD camera (Scout
scA640-74f, Basler).
Further, we apply the external magnetic fields 
using 
a set of custom-made magnetic coils mounted 
directly on the microscope stage. The rotating field in the $(\hat{\bm{x}},\hat{\bm{z}})$
plane is obtained by connecting
two set of coils to a wave generator (TTi-TGA1244, TTi) feeding a power
amplifier (IMG STA-800, stage line).

\section{Microroller propulsion}
To propel the hematite particles
we use a rotating magnetic field
circularly polarized in the $(\hat{\bm{x}},\hat{\bm{z}})$ plane, 
$\bm{H}=H(\cos{(2\pi f t)}\hat{\bm{x}}-\sin{(2 \pi f t)}\hat{\bm{z}})$, being $H$ the amplitude and $f$ the driving frequency, see Fig.\ref{figure1}(a).
As an example, in Fig.\ref{figure1}(c) we show two microscope snapshots of a collection of hematite rollers 
that are driven toward right by a rotating field with frequency $f = 100$Hz, reaching an average translational speed $\bar{v}_x=12.0 \pm	4 \, 
\rm{\mu m s^{-1}}$. 
We indicate with $v_i$, the single particle velocities in the plane 
$(\hat{\bm{x}},\hat{\bm{y}})$, and with $\bar{v}_i$ the collective speed, where $i=(x,y)$.
As the magnetic moments in each particle are directed along their short axis,
the particles roll around their long axis. Since we are interested in the collective particle transport, in this work we keep fix the field amplitude to $H=4400 \, \rm{A m^{-1}}$,
and vary mainly the  driving frequency $f$ and the particle density $\phi=N/A$, 
being $N$ the number of particles in the observation region $A$.  
Further, we note that above $\phi>0.04 \rm{\, \mu m^{-2}}$ the magnetic rollers start to overlap and propel also in the third dimension. This situation complicates the 
particle tracking, thus we limit our measurements to lower particle densities. 

Before exploring the collective transport, 
we start by showing in Fig.\ref{figure1}(d) how the single particle 
speed changes with $f$. 
For a rotating field polarized in the $(\hat{\bm{x}},\hat{\bm{z}})$ plane, the hematite particles display a negligible transverse velocity ($v_y \sim 0$)
while the translational motion along the $\hat{\bm{x}}$-axis can be characterized by two dynamic regimes.
Below a critical frequency  $f_c= 108.4$Hz, 
the magnetic
moment of the particle follows the
applied field. In this synchronous regime the particle rotational spinning coincides with the driving one, 
and the rolling velocity is $v_x  = 2\pi a \gamma f$
where $a=\beta/2$, and $\gamma$ is a prefactor that takes into account the close proximity of the wall.
The continuous blue line in Fig.\ref{figure1}(d) shows the corresponding fit to the experimental data in this linear regime, where we get $\gamma= (3.19 \pm 0.09) \cdot 10^{-2}$. In contrast, 
for $f>f_c$  the motion becomes asynchronous,
and this produces a reduction of the translational speed which 
can be described as $v_x  = 2\pi a \gamma (f-\sqrt{f^2-f_c^2})$,~\cite{Adler1946,Helgesen1990}
as shown by the continuous blue line in Fig.\ref{figure1}(d).
We note that, given the large density of the hematite particles ($\rho=5.3 \rm{g \, cm^{-3}}$ in the bulk), some rollers may become close to the surface especially at low spinning frequencies ($f<<f_c$ and $f>>f_c$), when the hydrodynamic flow generated by the spinning is low. In such cases, there can be some effects due to surface friction. We discuss the role of the different forces acting on the single propelling rotor in the Appendix of the manuscript.

\section{Collective transport.}
At large density, the hematite rollers start to interact 
due to their magnetic moments and the generated hydrodynamic flow.  
In agreement with previous observations for pair of particles,~\cite{Mar18} 
we find that close to $f_c$ the HIs dominate over magnetic one, and the particle preferentially assemble tip to tip. In this configuration magnetic dipolar interactions are repulsive, since the hematite moments rotate along parallel planes. This effect can be quantified if one consider the magnetic dipolar interaction $U_{m}$ between a pair of particles $(i,j)$, at positions $\bm{r}_i$ and $\bm{r}_j$ and with moments $\bm{m}_i$ and $\bm{m}_j$. Assuming point like particles, this interaction reads
 \begin{equation} 
U_{m}=\frac{\mu_w}{4\pi}\left( \frac{{\bm m}_i {\bm m}_j}{r_{ij}^3}-\frac{3({\bm m}_i \cdot {\bm r}_{ij})({\bm m}_j\cdot {\bm r}_{ij}) }{r_{ij}^5}\right)
\label{dipolar}
\end{equation}
were $\mu_w \sim \mu_0 = 4 \pi \cdot 10^{-7} \, \rm{H \cdot m}$
is the magnetic susceptibility of the medium (water), and $r_{ij}=|\bm{r}_i-\bm{r}_j|$
the separation distance. Eq.\ref{dipolar} shows that $U_{m}$
is maximally attractive (repulsive) for particles with
magnetic moments parallel (perpendicular) to $\bm{r}_{ij}$.
If one perform a time averaging, it is straightforward to show 
that hematite particles rolling side by side
are effectively attractive with $\overline{U}_m = -\mu_0 m^2/(8\pi (x+z)^3)$.
In contrast, when the particles are aligned tip to tip on the same plane, the moments are always perpendicular to their separation distance, and $U_m$ is repulsive, $\overline{U}_m = \mu_0 m^2/(4\pi y^3)$.  
%
\begin{figure}[t]
\includegraphics[width=0.9\columnwidth,keepaspectratio]{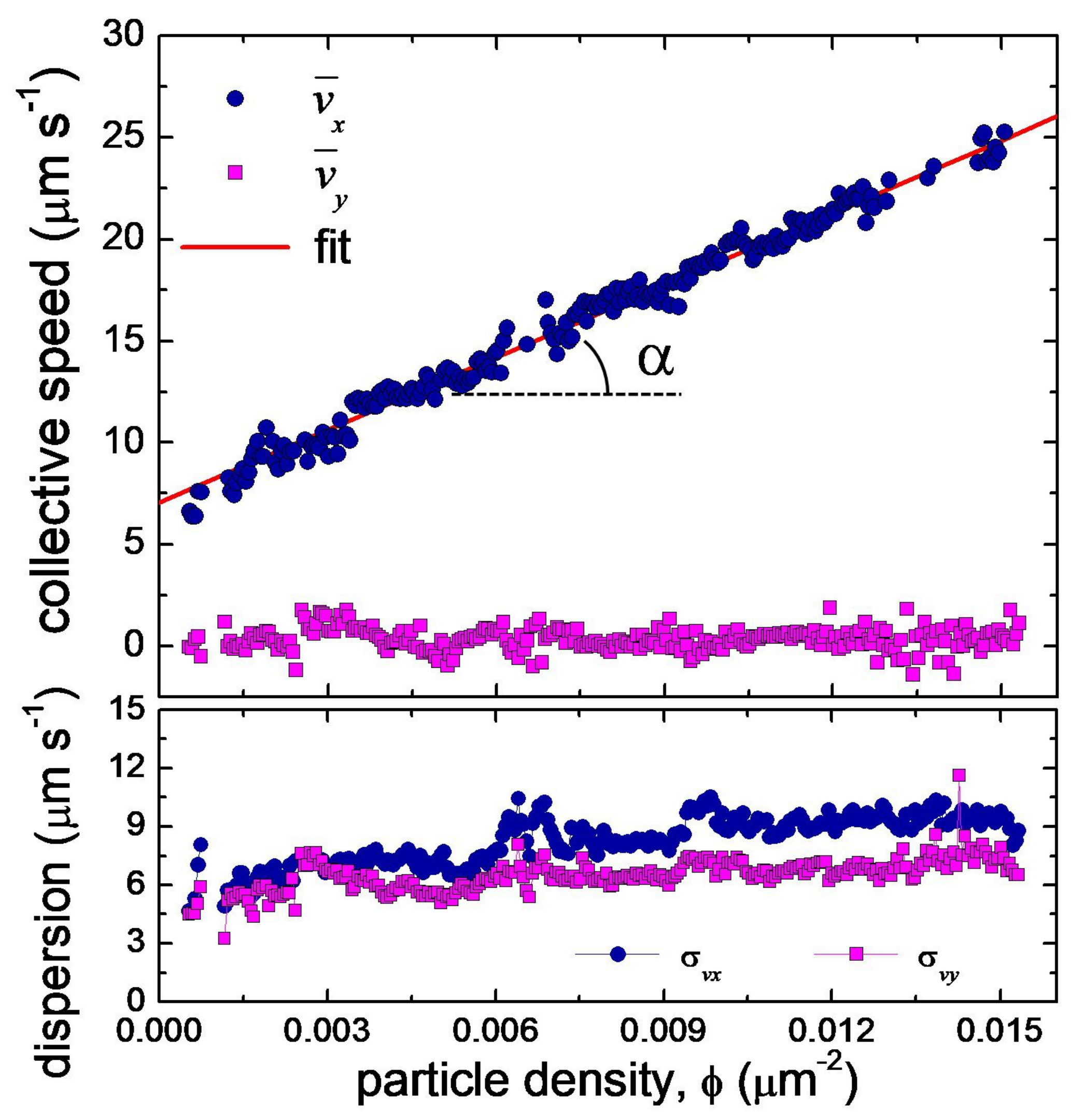}
\caption{ 
Top: collective mean velocities $\bar{v}_i$ 
with $i=(x,y)$ along ($\bar{v}_x$, blue disks) and perpendicular to ($\bar{v}_y$, magenta squares) the propulsion direction versus the roller's particle density $\phi$
for $f=100$Hz.
Scattered points are experimental data,
the continuous red line denotes 
the linear regression $\bar{v}_x=v_0+\alpha \phi$,
with $\alpha = 1183 \pm 12 \, \rm{\mu m^3 s^{-1}}$ and $v_0=7.08 \pm 	0.11 \, \rm{\mu m s^{-1}}$
Bottom:
corresponding velocity fluctuations $\sigma_{vx}$ and $\sigma_{vy}$.
\label{figure2}}
\end{figure}
%
From the experimental observations, we find that at low frequency $f <50 $Hz,
the particles tend usually to form compact clusters that grow along the propulsion direction, thus favouring chaining due to attractive dipolar force, see VideoS1 in the Supporting Information. 
However, for higher frequency these clusters 
form and break preferentially along the perpendicular direction, 
and thus their formation is promoted by HIs, being $U_m$ repulsive, see VideoS2.   

We now focus on the collective velocity as function of
the particle density.
As shown in Fig.~\ref{figure2}, $\bar{v}_x$  
linearly raises with the particle density $\phi$,
while being negligible the transverse velocity. 
The fluctuations in speed $\sigma_{vx,vy}$ also increase with $\phi$ along both directions. At high frequencies, when HIs dominates, we find that such interactions induce the formations of fragile clusters which grow perpendicular to the propulsion directions, see supporting videos. These clusters continuously break and reform, and this effect induces a raise of the speed fluctuations along the propulsion directions due to their fragmentation process, and perpendicular to it due to their anisotropic growth.

We explain this collective speed up effect
by considering only HIs and taking into account the proximity of the wall.
We assume an homogeneous distribution of point like particles
with density $\phi$,
each subjected to a magnetic torque $\bm{T}_m=(0,T_m,0)$.
For a single particle at position 
$(x',y',h)$ above a no-slip surface at $z=0$,
the translational velocity is given by 
\begin{equation}
v_{x}(x,y,z,x',y',h)=\frac{T_m}{8\pi\eta}\Bigl(\frac{z-h}{\Delta_{-}^{3/2}}-\frac{z-h}{\Delta_{+}^{3/2}}+\frac{6z(x-x')^{2}}{\Delta_{+}^{5/2}}\Bigr)
\label{Eq:1}
\end{equation}
where $\Delta_{\pm}=(x-x')^{2}+(y-y')^{2}+(z\pm h)^{2}$.
The velocity field created by the ensemble of rollers 
is obtained as:
\begin{equation}
\bar{v}_{x}(x,y,z)=\int^{\infty}_{-\infty}dx'\int^{\infty}_{-\infty}dy' \phi v_{x}(x,y,z,x',y',h)  \, \, . \nonumber
\end{equation} 
Carrying out this integration gives rise to a discontinuous velocity profile $\bar{v}_{x}=(T_m \phi)/(2\eta)$ for $z>h$ and $\bar{v}_{x}=0$ for $z< h$.
The obtained relationship shows that effectively the particle speed 
grows proportionally to the density of rollers in a two dimensional plane. We note that such density dependence was shown to give rise to the Burger equation for the particle concentration.~\cite{Belovs2014}

To apply this result to our hematite rollers, we derive the translational speed of the single particle subjected to the magnetic torque $\bm{T}$.
According to Ref.\cite{Goldman1967}, the linear $v_x$ and angular $\Omega$ velocities of a force-free spherical particle near a solid wall under  an external torque $T_m$ must satisfy the conditions
\begin{equation}
6\pi\eta a F^{*}_{t}v_x+6\pi\eta a^{2} F^{*}_{r}\Omega=0
\label{Eq5}
\end{equation}
and
\begin{equation}
T_m=8\pi\eta a^{2}T^{*}_{t}v_x+8\pi\eta a^{3}T^{*}_{r}\Omega
\label{Eq6}
\end{equation}
where we consider the following friction coefficients approximating rotating particles by spheres: \cite{Goldman1967} 
\begin{eqnarray}
F^{*}_{t}=-\frac{8}{15}\ln{(\delta)}+0.9588 \nonumber \\
F^{*}_{r}=\frac{2}{15}\ln{(\delta)}+0.2526 \nonumber \\
T^{*}_{r}=-\frac{2}{5}\ln{(\delta)}+0.3817 \nonumber \\
T^{*}_{t}=\frac{1}{10}\ln{(\delta)}+0.1895  \, . \nonumber \\
\end{eqnarray}
Here $\delta=h/a-1$ characterizes the thickness of the liquid layer between the particle and the no-slip surface. Eqs. (\ref{Eq5},\ref{Eq6}) give
\begin{equation}
v_{x}\equiv v_{0} =\frac{T_m}{8\pi\eta a^{2}}\frac{F^{*}_{r}}{F^{*}_{r}T^{*}_{t}-T^{*}_{r}F^{*}_{t}}
\end{equation}
and
\begin{equation}
\bar{v}_{x}/v_{0}=4\phi \frac{F^{*}_{r}T^{*}_{t}-T^{*}_{r}F^{*}_{t}}{F^{*}_{r}}  \, \, .
\label{Eq7}
\end{equation}
We use Eq.\ref{Eq7} to fit the experimental data in Fig.\ref{figure2}, where we keep as sole variable the particle elevation $\delta$ from the surface.
From the linear regression we determine the elevation $h=0.66\, \rm{\mu m}$ 
from the centre of the particle to the plane. Since the hematite particles roll perpendicular to their long axis, we obtain that the surface to surface elevation is given by $h-a = 200$nm, a reasonable value similar to that found in previous works with other magnetic colloids floating close to a plane in water.~\cite{Bli05,Pedrero2015} 
\begin{figure}[t]
\includegraphics[width=\columnwidth,keepaspectratio]{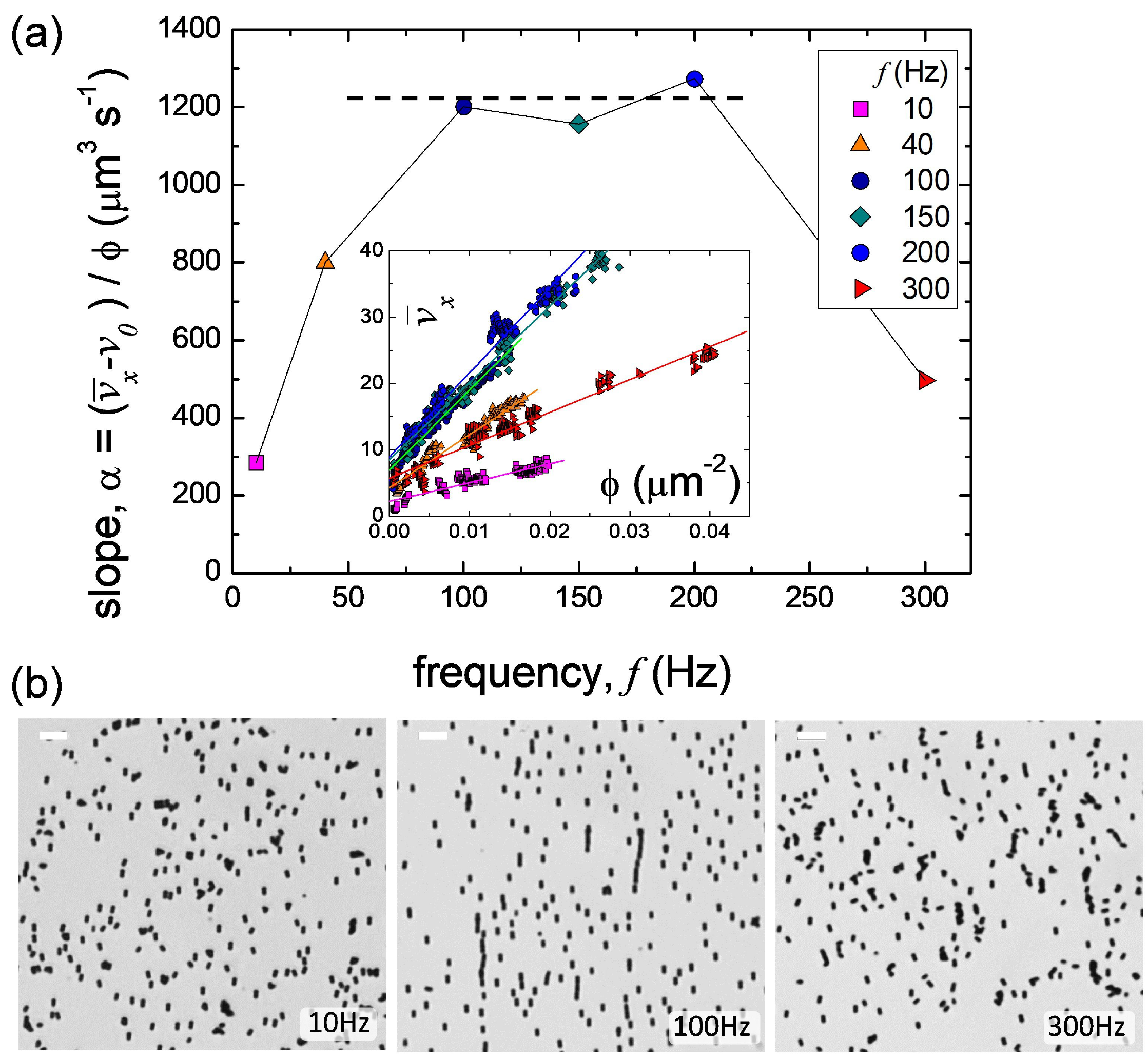}%
\caption{(a) Slope $\alpha$ of the collective velocity \textit{vs} driving frequency $f$. The dashed line is a guide to the eye. 
Inset shows the corresponding collective mean velocities $\bar{v}_x$ \textit{vs}  $\phi$ for different frequencies. Scattered points are experimental data, continuous lines are linear regressions.
(b) Three microscope images illustrating typical clusters 
observed at $f=10$Hz (left), $f=100$Hz (middle) and $f=300$Hz (right).
Scale bar is $10 \, \rm{\mu m}$ for all images. 
The corresponding videos can be found in the Supporting Information.
\label{figure3}}
\end{figure}
%
%
\begin{figure*}[t]
\includegraphics[width=\textwidth,keepaspectratio]{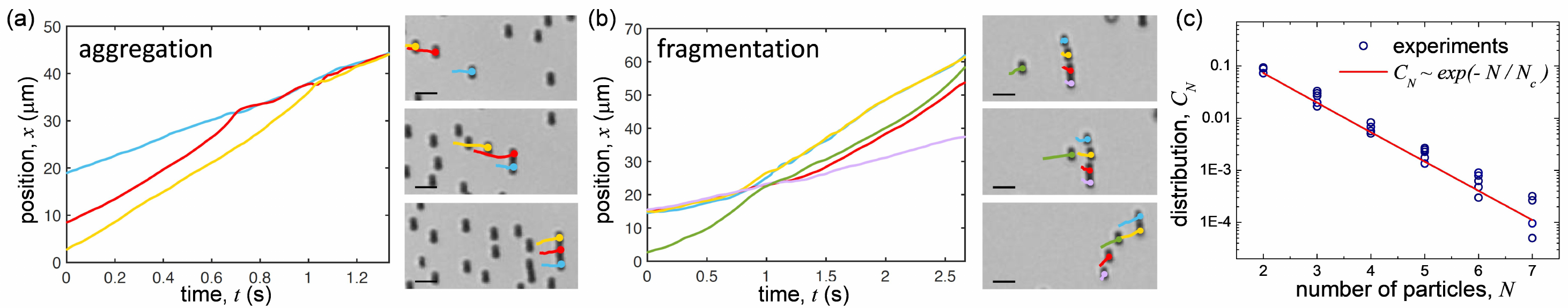}%
\caption{(a,b) Position $x$ versus time $t$ 
of different hematite rollers during the aggregation (a) and fragmentation (b) process of a cluster. The optical microscope images on the right 
with superimposed the particle trajectories over the last $0.25$s,
serve to illustrate the corresponding 
dynamic processes. Scale bars are $10\, \rm{\mu m}$ for all images.   
(c) Cluster size distribution $C_N$ versus number of particles $N$ 
at a constant particle density 
$\phi \sim 10^{-2} \, \rm{\mu m^{-2}}$. Empty symbols are different experiments while the continuous red line is an exponential fit. The particles are driven by a rotating field with amplitude $H=4400 \, \rm{A m^{-1}}$ and frequency $f=100$Hz.
\label{figure4}}
\end{figure*}
%
We note that even if our approach uses a mean-field like description, neglecting any correlation effect between the particles, it captures very well the experimental findings. Further, we note that
in the case of stress-free boundary at $z=0$ 
one can demonstrate that surface rollers under the same torque $\bm{T}$ 
move along the opposite direction. 
Indeed using the method of images \cite{Lauga2020} 
the velocity field is given by 
\begin{equation}
\bm{v}=\frac{\bm{T}\times\bm{r}}{8\pi\eta r^{3}}-\frac{\bm{T}\times\bm{r}^{*}}{8\pi\eta r^{*3}} \, \, ,
\label{Eq:3}
\end{equation}
where $\bm{r}^{*}=\bm{r}+2h\bm{z}$. In a similar way than Eq.(3), after integration we obtain $\bar{v}_{x}=0$ for $z>h$ and $\bar{v}_{x}=-(T \phi)/(2\eta)$ at $z<h$ satisfying the zero stress boundary condition at $z=0$. Thus, rollers located on the free interface move along the opposite direction with respect to the particles placed on a no-slip boundary. 
This conclusion was experimentally 
confirmed with chain of paramagnetic rotors.\cite{Pedrero2015}

The data shown in Fig.\ref{figure2} were obtained at $f=100$Hz, thus very close to $f_c$.
However, we also find that the mean velocity increases
with the density for other frequencies but, as $f<<f_c$ or $f>>f_c$, this effect
becomes less pronounced.
Fig.\ref{figure3}(a) shows the slope $\alpha = (\bar{v}_x-v_0)/\phi$ obtained from the linear regression for $6$ different frequencies, inset Fig.\ref{figure3}(a).
For $f=100,150$ and $200$ (dashed line) we find that $\alpha$ reaches a
maximum constant value, where HIs completely dominate over dipolar forces. In this situation, we observe the formation of clusters that grow perpendicular to the $\hat{\bm{x}}$-axis, as shown in the microscope image in the middle of Fig.\ref{figure3}(b). 
In contrast, at low or high frequencies, the slow particle spinning (Fig.\ref{figure1}(d))
reduces the effect of HIs favouring dipolar forces and thus the growth of clusters along the
propulsion direction  ($\hat{\bm{x}}$), see left and right images in Fig.\ref{figure3}(b).
Now attractive dipolar forces slow down the mean velocities of the clusters, reducing the speed up effect. This is consistent with other cases of clustering induced by dipolar interactions in one~\cite{Pedrero2015} and two dimensions~\cite{Martinez2015}
were the mean velocity of the rollers was observed to saturate with the number of particles. Further, the competition between dipolar interactions and HIs do not dissolve the clusters, but rather induce a precessional-like motion of the particles and the formation of more irregular structures, as shown in VideoS3 of the supporting information.

\section{Dynamic clustering}
At large density the magnetic rollers tend to aggregate in form of small dynamic clusters
which continuously break and reform during propulsion.  
Two typical situations are shown in Figs.\ref{figure4}(a,b).
The first one (Fig.\ref{figure4}(a)) illustrates the formation of a three particle cluster 
that assembles along the  ($\hat{\bm{y}}$)-direction. Once formed, the clusters usually propel at a similar speed than the composing particles. 
The lifetime of the clusters is relative short as interaction with other incoming particles can easily break them. This is illustrated in Fig.\ref{figure4}(b) where one particle is able of breaking a four particle cluster. During this fragmentation process, the hematite particles speed up, since they are literally pushed away by the 
flow field of the incoming roller. The microrollers form rather "fragile" hydrodynamic states, in contrast to  
other clusters made of active 
or self-propelling particle systems.~\cite{Theurkauff2012,Buttinoni2013,Julian2015,Massana2018,Ginot2018}
In Fig.\ref{figure4}(c) we analyse the cluster size 
distribution $C_N$ at a relative large density of particles, $\phi \sim 10^{-2} \, \rm{\mu m^{-2}}$. Even if the maximum cluster size observed is limited to $N=7$,
we find that the data follow a simple exponential law, $C_N \sim \exp{(-N/N_c)}$ which gives a characteristic value $N_c=0.78$. 
This behaviour is different than the one observed in other active systems,
where the dynamic clustering process is usually describe by a mixed exponential and power law function.~\cite{Ginot2018} 
In our case, the absence of an initial power law scaling in $C_N$
reflects the different cluster nature, where HIs make these structures rather fragile. 
Indeed, clusters made of chemically active particles~\cite{Theurkauff2012,Pohl2015} or bacteria~\cite{Chen2012,Petroff2015,Kirkegaard2016} can grow isotropically over a much larger area  
and the composing units present, apart from the random orientation 
during propulsion, other interactions competing with HIs. Instead our driven microrollers are characterized by 
almost deterministic trajectories with small thermal 
fluctuations, and the cluster aggregation/fragmentation process is 
driven by the sole HIs.

\section{Conclusions}
In this work we investigate the collective transport properties
of anisotropic hematite rollers 
that are driven in a viscous fluid by an external rotating magnetic field. 
We find that at high driving frequency, HIs 
dominate and induce the formation of elongated clusters 
perpendicular to the direction of motion, 
even if dipolar interactions are repulsive for such configuration. 
Moreover the particle collective mean velocity 
linearly increases with the density, a fact that we explain on the basis of 
Blake's tensor formalism applied to our colloidal rollers.~\cite{Bla74}

We note that 
Driscoll \textit{et al.}~\cite{Driscoll2017}
reported recently that a collection of microscopic magnetic rollers made of spherical particles with protruding hematite cubes, display a fingering instability produced by the hydrodynamic flow
generated by their spinning. 
In such work the authors  also report that the velocity of the rollers linearly grow with the particle density, and the density-dependent mean velocity has been later used 
to predict the formation of shock fronts in this system.~\cite{Delmotte2017} 
Our hematite colloidal rollers display relative stronger magnetic forces, 
a feature that may be used to further control
dipolar interactions in contrast to hydrodynamic one. 
Indeed, it would be interesting to investigate in the future weather HIs  can play an important role in the synchronization 
process of the magnetic rollers with the rotating field. 

\section{Appendix}
\subsection{Single particle dynamics}
Isolated hematite particles are subjected to different forces during motion. Given the large density mismatch, $\Delta \rho = \rho_p-\rho_w$, between hematite ($\rho_p=5.3 \, \rm{g cm^{-3}}$ in bulk) and water ($\rho_w=1 \, \rm{g cm^{-3}}$), the gravitational force acting on a spherical particle of radius $a=\beta/2$ can be estimated as $F_g =  \frac{4}{3}\pi  a^3 \Delta \rho g \sim 0.06 \rm{pN}$, thus much higher than lift force $F_l \sim 60 \rm{fN}$, calculated~\cite{Rashidi2016,Helgessen2018} for a similar spherical particle spinning at the maximum frequency of $f=100$Hz.
Thus, in absence of repulsive electrostatic interactions that float the particle at a distance of $\sim 200$nm from the surface, the particles would touch the solid surface and display a stick-slip dynamics.

Another effect which may affect the roller dynamics at very low and high frequency, is the surface roughness. Its effect may be present in the asynchronous regime (see VideoS3 of the Supporting Information) when the particle perform a wobbling-like movement. 
We consider the particle roughness \cite{Smart1989} in terms of the typical size of surface asperities or bumps, $\varepsilon a$. The solid friction becomes important when the size of the surface bumps $\varepsilon a$ is of the order of the thickness of the gap between the particle and plane $(h-a)$ \cite{Smart1993}. 
In this situation, the hydrodynamic problem for a particle moving near the plane \cite{Goldman1967} is given by the following force and torque balance equations
\begin{equation}
F_s-6\pi\eta a F^{*}_{t}v_{x}-6\pi\eta a^{2}F^{*}_{r}\Omega=0
\label{Eq:1}
\end{equation}
and
\begin{equation}
T_{m}+F_s a-8\pi\eta a^{3}T^{*}_{r}\Omega-8\pi\eta a^{2}T^{*}_{t}v_{x}=0
\label{Eq:2}
\end{equation}
where $F_s$ is the solid friction force and $T_{m}$ is the torque from the rotating magnetic field. Two different situations may be considered - motion without slip when $v_{x}=a\Omega$ and $F_s<F_s^{max}=kN$, where $k$ is a solid friction coefficient and $N=F_g$ the normal force. A simple estimate for the hematite particle under a rotating field with amplitude $H=4000 \rm{A m^{-1}}$ of the ratio $\tilde{T_{m}}=MHV/(\Delta\rho g V a)$ shows that $\tilde{T_{m}}>>1$ and the particle is in slipping regime when $F_s=F_s^{max}$. Scaling $v_{x}$ with the sedimentation velocity $N/(6\pi\eta a)$ and the angular velocity by $N/(6\pi\eta a^{2})$ give
\begin{equation}
F^{*}_{t}\tilde{v_{x}}+F^{*}_{r}\tilde{\Omega}=k
\label{Eq:3}
\end{equation}
and
\begin{equation}
\tilde{T_{m}}=\frac{4}{3}T^{*}_{r}\tilde{\Omega}+\frac{4}{3}T^{*}_{t}\tilde{v_{x}}-F^{*}_{t}\tilde{v_{x}}-F^{*}_{r}\tilde{\Omega}
\label{Eq:4}
\end{equation}
From Eq.(\ref{Eq:3})
\begin{equation}
\tilde{\Omega}=\frac{k}{F^{*}_{r}}-\frac{F^{*}_{t}\tilde{v_{x}}}{f^{*}_{r}}
\label{Eq:5}
\end{equation}
Inserting that in Eq.(\ref{Eq:4}) we have
\begin{equation}
\tilde{v_{x}}=\frac{3}{4}\tilde{T_{m}}\frac{F^{*}_{r}}{F^{*}_{r}T^{*}_{t}-T^{*}_{r}F^{*}_{t}+k(T^{*}_{r}-\frac{3}{4}F^{*}_{r})}
\label{Eq:6}
\end{equation}
In dimensional units this gives
\begin{equation}
v_{x}=\frac{T_{m}}{8\pi\eta a^{2}}\frac{F^{*}_{r}}{F^{*}_{r}T^{*}_{t}-T^{*}_{r}F^{*}_{t}+k(T_{r}^{*}-\frac{3}{4}F_{r})}
\label{Eq:7}
\end{equation}
As a result for the ratio of the collective velocity of ensemble $v_{c}=T_{m}\phi/(2\eta)$ and velocity of the single particle $v_{x}$ we have
\begin{equation}
\frac{v_{c}}{v_{x}}=4\varphi\frac{F^{*}_{r}}{F^{*}_{r}T^{*}_{t}-T^{*}_{r}F^{*}_{t}+k(T_{r}^{*}-\frac{3}{4}F_{r})}
\label{Eq:8}
\end{equation}
where $\varphi=\pi a^{2}\phi$ is the surface fraction covered by the magnetic particles.

\section{Acknowledgments}
G. J. and P. T. acknowledge support from the European Research Council (ERC) under the European Union’s
Horizon 2020 research and innovation programme (grant agreement No.: 811234).
A.C. acknowledge support by grant of Scientific Council of Latvia lzp-2020/1-0149.
P. T. acknowledges support 
from MICIU (grant No.  PID2019-108842GB-C21),
AGAUR (grant No. 2017-SGR-1061) and from the 
Generalitat de Catalunya via program "Icrea Acad\`emia".

\footnotesize{
\providecommand*{\mcitethebibliography}{\thebibliography}
\csname @ifundefined\endcsname{endmcitethebibliography}
{\let\endmcitethebibliography\endthebibliography}{}

\bibliographystyle{rsc} 
}

\end{document}